\documentclass[cameraready]{Interspeech}

\title{WhispEar: A Bi-directional Framework for Scaling Whispered Speech Conversion via Pseudo-Parallel Whisper Generation}

\author[affiliation={1}, orcid=0009-0003-0161-7176, equalcontribution]{Zihao}{Fang}
\author[affiliation={1}, orcid=0009-0003-6718-7180, equalcontribution]{Yingda}{Shen}
\author[affiliation={1}, orcid=0009-0006-6830-028X]{Zifan}{Guan}
\author[affiliation={2}]{Tongtong}{Song}
\author[affiliation={2}]{Zhenyi}{Liu}
\author[affiliation={1}, orcid=0009-0001-1192-9857]{Zhizheng}{Wu}

\address{
    $^1$ School of Data Science, The Chinese University of Hong Kong, Shenzhen, China \\
    $^2$ Honor Device Co., Ltd, China
}

\email{zihaofang@link.cuhk.edu.cn, yingdashen@link.cuhk.edu.cn}


\usepackage{comment}
\usepackage[table,xcdraw]{xcolor}
\usepackage{multirow}
\usepackage{amsmath, amssymb, amsfonts} 
\usepackage{bm}                           
\usepackage{graphicx}
\usepackage{float}

\begin{document}

\maketitle

\begin{abstract}
Whispered speech lacks vocal fold vibration and fundamental frequency, resulting in degraded acoustic cues and making whisper-to-normal (W2N) conversion challenging, especially with limited parallel data. We propose WhispEar, a bidirectional framework based on unified semantic representations that capture speaking-mode-invariant information shared by whispered and normal speech. The framework contains both W2N and normal-to-whisper (N2W) models. Notably, the N2W model enables zero-shot pseudo-parallel whisper generation from abundant normal speech, allowing scalable data augmentation for W2N training. Increasing generated data consistently improves performance. We also release the largest bilingual (Chinese–English) whispered–normal parallel corpus to date. Experiments demonstrate that WhispEar outperforms strong baselines and benefits significantly from scalable pseudo-parallel data.
\end{abstract}

\renewcommand{\thefootnote}{\fnsymbol{footnote}}
\footnotetext[0]{The demo page is accessible at \href{https://whispear-demo.github.io/}{https://whispear-demo.github.io/}}
\renewcommand{\thefootnote}{\arabic{footnote}}

\section{Introduction}

Whisper-to-Normal conversion aims to reconstruct natural and intelligible speech from whispered speech, which is crucial for privacy communication and voice restoration. Due to the absence of vocal fold vibration and periodic excitation, whispered speech suffers from the loss of fundamental frequency and degraded acoustic characteristics\cite{tartter1989s}, bringing great challenges to generating speech with natural prosody and consistent timbre. Although existing studies have promoted the development of W2N conversion, they still face severe limitations. On the one hand, these approaches heavily rely on limited parallel whispered-normal data\cite{seq2seq,agan-w2n}, and conventional DSP-based pseudo-whisper data\cite{normal2whisper,toWhisper} exhibits a distribution gap with real whispered speech, providing limited performance gains. Adversarial learning-based methods further suffer from training instability\cite{disco&cyclegan,voc-free-cyclegan}. On the other hand, most methods struggle to preserve speaker timbre and natural prosody\cite{wesper,parrotron}, leading to unsatisfactory generation quality and low speaker similarity.

Content disentanglement-based models have achieved remarkable success in text-to-speech and voice conversion\cite{du2024cosyvoice, zhang2025vevo}, benefiting from their strong ability to generate high-quality speech while maintaining timbre and prosody. Inspired by this, recent W2N methods such as WESPER\cite{wesper} and DistillW2N\cite{distillw2n} have introduced semantic representations like HuBERT\cite{hsu2021hubert} to improve conversion performance.
In this work, we observe that while whispered and normal speech differ acoustically, they share the same high-level linguistic and semantic information. Based on this, both W2N and N2W tasks can be regarded as generation tasks that produce different styles from the same semantic content, which logically supports bidirectional conversion between the two speaking modes. 

To address the data scarcity issue and further improve W2N performance, we propose a semantic feature-based bidirectional conversion framework, \textbf{WhispEar}. Leveraging its N2W module, we can synthesize high-quality pseudo-whisper data from massive normal speech dataset\cite{haorui2024emilia}. In addition, we recruit volunteers to collect real whispered speech data. By combining real recorded data and high-quality pseudo data, we build a most large-scale bilingual whispered-normal dataset in this area to date, wEar. Experimental results demonstrate that our method effectively alleviates data limitations, and significantly outperforms baseline models in terms of speech naturalness, intelligibility, prosody recovery and speaker timbre similarity.

The main contributions of this paper are:
\begin{itemize}

    \item We propose \textbf{WhispEar}, a bidirectional whispered speech conversion framework based on unified semantic representations. 

    \item We introduce a \textbf{pseudo-parallel whisper generation strategy} via zero-shot normal-to-whisper synthesis, allowing large-scale data expansion from abundant normal speech without additional recording effort.

    \item We conduct a \textbf{systematic scaling study} by progressively increasing generated pseudo-parallel data, demonstrating consistent performance gains and validating the effectiveness of data-centric scaling for whispered speech conversion.

    \item We release the \textbf{largest bilingual (Chinese and English) whispered-normal parallel corpus} to date, including both recorded and generated data, providing a valuable benchmark for future research.

\end{itemize}

\begin{figure*}[t]
    \centering
    \includegraphics[width=0.7\textwidth]{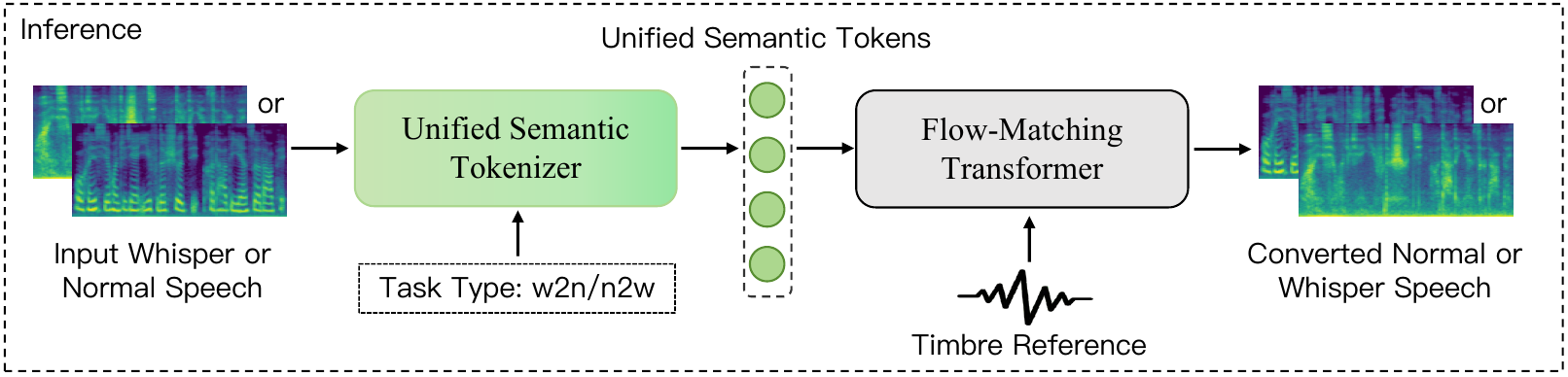}
    \caption{Inference pipeline of WhispEar for bidirectional conversion.}
    \label{fig:inference}
    \vspace{-8pt}
\end{figure*}

\begin{figure}[t]
    \centering
    \includegraphics[width=\columnwidth]{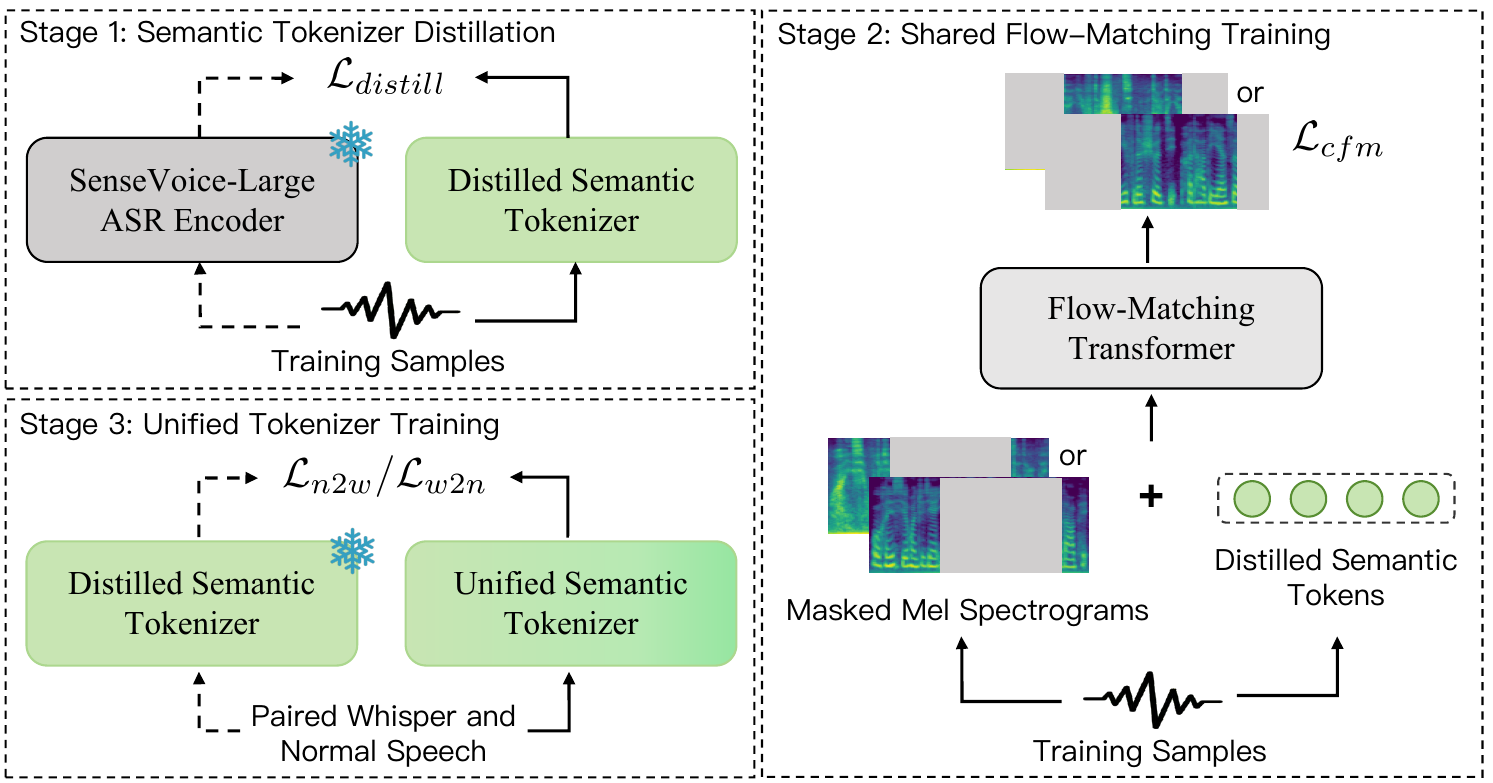}
    \caption{Training pipeline of WhispEar.}
    \label{fig:training}
    \vspace{-8pt}
\end{figure}

\section{Methodology}

WhispEar is trained in three sequential stages:
(1) semantic tokenizer distillation,
(2) shared flow-matching acoustic model training, and
(3) unified tokenizer training with scalable pseudo-parallel data generation.

\subsection{Stage 1: Semantic Tokenizer Distillation}

We first distill a lightweight semantic tokenizer from a large-scale ASR encoder.
Given waveform $x$, the teacher model $f_{\text{large}}$ produces semantic embeddings.
A compact student model $f_\phi$ is trained to mimic the teacher via

\begin{equation}
\mathcal{L}_{\text{distill}}
=
\mathbb{E}_{x}
\left\lVert
f_\phi(x) - f_{\text{large}}(x)
\right\rVert_2^2
.
\end{equation}

The student tokenizer consists of stacked Transformer blocks, each composed of RoPE-based self-attention, an FSMN block, and a feed-forward network (FFN).
The FSMN (Feedforward Sequential Memory Network) is a lightweight temporal modeling module that captures local sequential dependencies using a convolutional memory mechanism without recurrent connections, providing efficient contextual smoothing.

Training data include both whispered and normal speech, encouraging speaking-mode-invariant representations.
The resulting embeddings are quantized using Finite Scalar Quantization (FSQ)~\cite{mentzer2024fsq} to obtain discrete semantic tokens.
These tokens are used to train the acoustic model in Stage 2.

\subsection{Stage 2: Shared Flow-Matching Acoustic Model}

We train a conditional Flow-Matching Transformer to generate mel-spectrograms from discrete semantic tokens.
Both whisper-to-normal (w2n) and normal-to-whisper (n2w) share the same Flow-Matching Transformer and the same vocoder.
A direction indicator $d \in \{\text{w2n}, \text{n2w}\}$ specifies the conversion direction.

Importantly, the acoustic model is trained on tokens produced by the \emph{distilled} tokenizer $f_\phi$, not the unified tokenizer introduced later.
This decouples acoustic modeling from semantic alignment.

Following CosyVoice2, the model takes as input the semantic tokens $\mathbf{Z}$ together with a masked mel-spectrogram.
The masked regions are initialized with Gaussian noise, and the model predicts the velocity field of the masked portion along the optimal-transport probability path.

The training objective is
\begin{equation}
\mathcal{L}_{\text{cfm}}
=
\mathbb{E}
\left\lVert
(\mathbf{Y}_1 - \mathbf{Y}_0)
-
\nu_t(\mathbf{Y}_t \mid \mathbf{Z}, d)
\right\rVert_1
.
\end{equation}

where the loss is computed only on masked regions.
The model is initialized from CosyVoice2 and fine-tuned on mixed whispered and normal speech.

\subsection{Stage 3: Unified Tokenizer Training and Scalable Expansion}

After fixing the acoustic model, we train two unified semantic tokenizers:
$f_{\text{n2w}}$ and $f_{\text{w2n}}$.
These models map speech from one modality into the target semantic token space of the other.

Let $\mathcal{T}(\cdot)$ denote the frozen distilled tokenizer.
For paired whispered-normal speech $(x_w, x_n)$, we extract
$\mathbf{z}_w = \mathcal{T}(x_w)$ and
$\mathbf{z}_n = \mathcal{T}(x_n)$.

The training objectives are:
\begin{equation}
\begin{aligned}
\mathcal{L}_{\text{w2n}}
=
\mathbb{E}_{(x_w,x_n)\sim\mathcal{P}}
\Big[
&\left\lVert
f_{\text{w2n}}(x_w) - \mathbf{z}_n
\right\rVert^2 \\
&+
\lambda_n
\left\lVert
f_{\text{w2n}}(x_n) - \mathbf{z}_n
\right\rVert^2
\Big].
\end{aligned}
\end{equation}
\begin{equation}
\begin{aligned}
\mathcal{L}_{\text{n2w}}
=
\mathbb{E}_{(x_w,x_n)\sim\mathcal{P}}
\Big[
&\left\lVert
f_{\text{n2w}}(x_n) - \mathbf{z}_w
\right\rVert^2 \\
&+
\lambda_w
\left\lVert
f_{\text{n2w}}(x_w) - \mathbf{z}_w
\right\rVert^2
\Big].
\end{aligned}
\end{equation}

\textbf{Step 1: Train the easier n2w direction:} Normal-to-whisper conversion is empirically easier and achieves good quality with limited paired data.
We first train $f_{\text{n2w}}$ using real paired data only.

\textbf{Step 2: Generate pseudo-parallel data:} Using abundant normal speech corpora and the trained n2w pipeline, we synthesize whispered speech $\tilde{x}_w$ from real normal speech $x_n$.
This yields fully aligned pseudo pairs $(\tilde{x}_w, x_n)$ at large scale.

\textbf{Step 3: Train the harder w2n direction with scaled data:} Finally, we train $f_{\text{w2n}}$ using both real paired data and large-scale pseudo-parallel data.
The expanded dataset significantly improves whisper-to-normal performance and enables systematic scaling experiments.

During inference, an input utterance is first mapped to target semantic tokens via the corresponding unified tokenizer, and the shared Flow-Matching Transformer and vocoder generate the output waveform.

\section{Dataset}
\label{sec:dataset}

To address the scarcity of large-scale bilingual whispered-normal corpora, we construct \textbf{wEar}, a parallel dataset covering both Chinese and English.

\subsection{Data Collection}

The wEar(Real) corpus was recorded in an empty classroom and a soundproof booth, including 146 speakers divided into parallel and whisper-only groups. Each speaker recorded 20--40 minutes of digits and emotionally varied utterances. Recordings were captured with three mobile phones at 48\,kHz/16-bit mono under background noise below 30\,dB, with speakers holding the phone near the ear to simulate phone calls. The dataset is split into training, validation, and test sets (91:6:3) with no speaker overlap.

\begin{table}[t]
\centering
\footnotesize
\renewcommand{\arraystretch}{1.2}
\setlength{\tabcolsep}{4pt}
\caption{Comparison of whispered speech conversion datasets in terms of language coverage, total duration, number of parallel pairs, and number of speakers.}
\label{tab:dataset}
\begin{tabular}{lcccc}
\hline
\textbf{Dataset} & \textbf{Lang.} & \textbf{Time(h)} & \textbf{Pairs} & \textbf{Spk} \\ 
\hline
\multicolumn{5}{c}{\cellcolor[HTML]{ECF4FF}\textit{Others}} \\ 
\hline
wTIMIT\cite{wtimit} & EN & 26 & 19k & 48 \\
CHAINs\cite{cummins2006chains} & EN & 3 & 1k & 36 \\
iWhisper\cite{lee2014whispered} & CN & 15 & -- & 80 \\
wSPIRE\cite{singhal2021wspire} & EN & 18 & -- & 88 \\
AISHELL6\cite{li2025aishell6} & CN & 30 & 20k & 167 \\
Whisper40\cite{yang2024whisper40} & CN & 6 & 1k & 36 \\
\hline
\multicolumn{5}{c}{\cellcolor[HTML]{ECF4FF}\textit{Ours}} \\ 
\hline
wEar (Real) & CN & 18 & 4k & 146 \\
wEar (Pseudo) & EN,CN & 3026 & 600k & 230 \\
\textbf{wEar (Total)} & \textbf{EN,CN} & \textbf{3044} & \textbf{604k} & \textbf{230} \\
\hline
\end{tabular}
\vspace{-6pt}
\end{table}

\subsection{Data Processing}

Waveforms are resampled and normalized, then processed by HuBERT ASR X-Large~\cite{hsu2021hubert} to obtain representations and spectrograms. After log-softmax and CTC greedy decoding, forced alignment determines frame boundaries, followed by word merging and silence trimming. FastDTW~\cite{salvador2007fastdtw} computes the alignment path, yielding quasi one-to-one frame mappings. Aligned spectrogram frames are then inverted to reconstruct time-domain signals, combining semantic-level alignment with frame-level precision.

\subsection{Data Construction}

To expand the corpus, we generate large-scale pseudo-parallel data using the proposed WhispEar model. For each speaker, high-quality speech clips are used as timbre prompts for voice conversion on utterances sampled from Emilia~\cite{haorui2024emilia}, producing timbre-matched speech $A_1$. The normal-to-whisper module (n2w) then generates whispered speech $A_2$, forming synthetic pairs $(A_1, A_2)$. This pipeline yields over 3{,}000 hours of pseudo-parallel data, substantially enlarging existing whisper--normal corpora and improving bidirectional conversion. Table~\ref{tab:dataset} summarizes the wEar dataset.

\section{Experiments}

Experiments are conducted to answer these evaluation questions: \textbf{EQ1:} How well does the WhispEar system perform in the whisper-to-normal (W2N) task compared to existing state-of-the-art methods? \textbf{EQ2:} How effective is our pseudo-whispered data, generated by WhispEar itself, for training a high-quality W2N system? \textbf{EQ3:} Can scaling up the pseudo-whispered data further improve the performance of the W2N system, and what is the role of supervised fine-tuning on real aligned data?

\begin{table*}[t]
\centering
\renewcommand{\arraystretch}{1.2}
\caption{Comparison with baselines on WTIMIT (English) and wEar (Chinese) testsets}
\label{tab:merged-baseline}
\resizebox{\textwidth}{!}{
\begin{tabular}{lcccccc|cccccc}
\hline
\multicolumn{1}{l|}{\textbf{Model}} 
& \multicolumn{6}{c|}{\textbf{WTIMIT (English)}} 
& \multicolumn{6}{c}{\textbf{wEar (Chinese)}} \\ \hline

& \textbf{SIM} & \textbf{WER} & \textbf{UTMOS} & \textbf{DNSMOS} & \textbf{NISQA} & \textbf{F0\_CoRR}
& \textbf{SIM} & \textbf{CER} & \textbf{UTMOS} & \textbf{DNSMOS} & \textbf{NISQA} & \textbf{F0\_CoRR} \\ \hline

\multicolumn{1}{l|}{Whispered Speech} 
& 0.333 & 12.36\% & 1.46 & 1.81 & 1.66 & 0.029
& 0.550 & 18.09\% & 1.41 & 1.84 & 1.59 & 0.016 \\ \hline

\multicolumn{13}{c}{\cellcolor[HTML]{ECF4FF}\textit{Baselines}} \\ \hline

\multicolumn{1}{l|}{WESPER\cite{wesper}} 
& 0.064 & 42.01\% & 3.63 & \textbf{3.91} & 4.37 & 0.373
& 0.004 & 141.36\% & \textbf{3.64} & \textbf{4.05} & \textbf{4.67} & 0.254 \\

\multicolumn{1}{l|}{DistillW2N\cite{distillw2n}} 
& 0.189 & 50.99\% & 1.97 & 3.61 & 2.25 & 0.113
& 0.216 & 128.06\% & 1.69 & 3.43 & 1.84 & 0.125 \\

\multicolumn{1}{l|}{MaskCycleGAN\cite{voc-free-cyclegan}} 
& 0.237 & 48.53\% & 1.97 & 3.24 & 3.32 & 0.023
& 0.322 & 49.06\% & 1.67 & 3.32 & 2.98 & 0.023 \\

\multicolumn{1}{l|}{CosyVoice2\cite{du2024cosyvoice}} 
& 0.409 & 36.69\% & 2.51 & 2.54 & 3.59 & 0.052
& 0.622 & 29.29\% & 2.11 & 2.59 & 3.20 & 0.062 \\ \hline

\multicolumn{13}{c}{\cellcolor[HTML]{ECF4FF}\textit{Ours}} \\ \hline

\multicolumn{1}{l|}{WhispEar} 
& 0.554 & 30.74\% & 3.45 & 3.70 & 4.15 & 0.485
& 0.702 & 32.50\% & 2.59 & 3.51 & 3.78 & 0.259 \\

\multicolumn{1}{l|}{WhispEar-Scaled} 
& \textbf{0.577} & \textbf{22.44\%} & \textbf{3.75} & 3.76 & \textbf{4.38} & \textbf{0.513}
& \textbf{0.750} & \textbf{14.93\%} & 3.16 & 3.94 & 4.32 & \textbf{0.396} \\ \hline

\end{tabular}}
\vspace{-6pt}
\end{table*}

\subsection{Experimental Setup}

\subsubsection{Training Data}

The training process leverages real aligned data and pseudo-whispered data generated by WhispEar-n2w model. The real aligned data consists of the wTIMIT train set (English) and the wEar train set (Chinese), which provide ground-truth whispered speech and corresponding normal speech after the alignment pipeline (see Section~\ref{sec:dataset}). To scale the training data and improve model generalization, large-scale pseudo-parallel data is generated via Whispear-n2w, covering pseudo versions of wTIMIT, wEar and a 3,000-hour pseudo Emilia subset. The statistics of all datasets are summarized in Table~\ref{tab:dataset}.

\subsubsection{Training Details}

We train two models: \textbf{WhispEar} and \textbf{WhispEar-Scaled}.
Both share Stage 1 and Stage 2, and differ only in Stage 3.

Stage 1: Distillation.
A student semantic tokenizer is distilled from a large ASR encoder using mixed whispered and normal speech.
No paired data are required.
The model is optimized with Adam at a learning rate of $1\times10^{-4}$.

Stage 2: Flow-Matching Training.
The acoustic model is trained using tokens from the distilled tokenizer.
Training settings follow CosyVoice2, except that we use a fixed learning rate of $1\times10^{-5}$ with Adam.
Mixed whispered and normal speech are used without requiring alignment.

Stage 3: Unified Tokenizer Training.

WhispEar:
Unified tokenizers are trained using aligned and pseudo-parallel data from wTIMIT and wEar.
Pseudo pairs are generated via the normal-to-whisper pipeline.
The total data size is approximately 80 hours.

WhispEar-Scaled:
We further incorporate large-scale pseudo-parallel data (approx. 3000 hours) generated from normal speech corpora to train the whisper-to-normal tokenizer.

\subsubsection{Evaluation Metrics}

We evaluate performance from four aspects: quality, intelligibility, prosody, and speaker similarity. 
Quality is measured by UTMOS~\cite{saeki2022utmos}, DNSMOS~\cite{reddy2021dnsmos}, and NISQA~\cite{mittag2021nisqa}. 
Intelligibility is assessed by WER (English) and CER (Chinese) using Whisper-large-v3~\cite{radford2022whisper}. 
Prosody is evaluated via the F0 Pearson correlation (F0\_CoRR) with target normal speech, and speaker similarity is computed as the cosine similarity between speaker embeddings extracted by wavlm-base-plus-sv~\cite{wavlm}.

\subsubsection{Baseline Models}

We compare with four representative W2N systems: 
WESPER~\cite{wesper}, 
DistillW2N~\cite{distillw2n}, 
CosyVoice2~\cite{du2024cosyvoice}, 
and MaskCycleGAN~\cite{voc-free-cyclegan}. 
The first three baselines are evaluated using their publicly released checkpoints, while MaskCycleGAN is re-trained using the official implementation on wTIMIT and wEar.

\subsection{Performance of WhispEar (EQ1)}

Table~\ref{tab:merged-baseline} reports the main results, including both \textbf{WhispEar} and \textbf{WhispEar-Scaled}.

In English benchmarks, WESPER achieves reasonable audio quality but suffers from low speaker similarity due to its single-speaker design. DistillW2N reduces model size via knowledge distillation but shows degraded intelligibility. CosyVoice2, optimized for normal speech, performs strongly in quality and similarity; however, its prosody recovery in W2N is weaker, reflected by lower F0 correlation.

\textbf{WhispEar}, trained without large-scale scaling (approx. 80h), already achieves competitive or superior performance compared with state-of-the-art systems across multiple metrics. 
After incorporating large-scale pseudo-parallel data (approx. 3000h), \textbf{WhispEar-Scaled} further improves consistently in similarity, intelligibility, naturalness, and prosody, achieving the best overall performance by a clear margin.

In wEar test set (Chinese), WESPER and DistillW2N (trained only in English) fail to generalize effectively, leading to severe intelligibility degradation ($\text{CER}>80\%$). In contrast, WhispEar maintains strong cross-lingual performance, and WhispEar-Scaled provides additional gains, showing the effectiveness of multilingual and scalable pseudo-data training.

\begin{table}[t]
\centering
\renewcommand{\arraystretch}{1.2}
\footnotesize
\setlength{\tabcolsep}{5pt}
\caption{Ablation study on different whispered data construction strategies using wTIMIT only, including raw pairs, DSP-generated whispers, aligned real pairs, WhispEar-generated pseudo pairs, and their combination.}
\label{tab:ablation}
\begin{tabular}{lcccc}
\hline
\textbf{Config} & \textbf{SIM} & \textbf{WER} & \textbf{UTMOS} & \textbf{F0\_CoRR} \\
\hline
RAW & 0.031 & 102.934\% & 1.326 & 0.009 \\
\hline
\multicolumn{5}{c}{\cellcolor[HTML]{ECF4FF}\textit{Traditional}} \\
\hline
DSP\cite{toWhisper} & 0.429 & 54.814\% & 2.446 & 0.208 \\
\hline
\multicolumn{5}{c}{\cellcolor[HTML]{ECF4FF}\textit{Proposed}} \\
\hline
Aligned (A) & 0.416 & 44.594\% & 1.598 & 0.309 \\
Pseudo (P) & 0.454 & 49.179\% & 2.595 & 0.271 \\
A + P & \textbf{0.513} & \textbf{34.871\%} & \textbf{3.064} & \textbf{0.357} \\
\hline
\end{tabular}
\vspace{-6pt}
\end{table}

\subsection{Effectiveness of Pseudo-whispered Data (EQ2)}
\label{sec:ablation}

Table~\ref{tab:ablation} presents an ablation study under different data configurations. 
All training data are derived exclusively from the original wTIMIT corpus; no external data are introduced. 
The configurations differ only in how the whispered speech is constructed and processed:

\begin{itemize}
    \item \textbf{RAW:} Directly uses the original normal and whispered recordings from wTIMIT without any alignment, where pairs are simply padded to equal length. Severe temporal mismatch leads to the worst performance.
    
    \item \textbf{DSP:} Uses the original normal recordings from wTIMIT, while whispered speech is generated from these normal signals via a traditional DSP-based normal-to-whisper method\cite{toWhisper}. Although temporally matched, the signal-processing conversion introduces a noticeable quality gap.
    
    \item \textbf{Aligned (A):} Uses the original normal and whispered recordings from wTIMIT, but applies the proposed alignment pipeline to obtain temporally aligned parallel pairs.
    
    \item \textbf{Pseudo (P):} Uses the original normal recordings from wTIMIT, while whispered speech is generated by the WhispEar N2W model to form pseudo-parallel pairs.
    
    \item \textbf{A + P:} Combines aligned real pairs (A) and WhispEar-generated pseudo pairs (P).
\end{itemize}

Results show that naive use of raw data severely degrades performance. Traditional DSP-based generation improves over RAW but remains limited. In contrast, both alignment (A) and WhispEar-generated pseudo data (P) provide substantial gains, and their combination (A+P) achieves the best results across all metrics. These findings demonstrate that high-quality temporal alignment and model-based pseudo-whisper generation are both critical for alleviating paired data scarcity.

\begin{figure}[h]
    \centering
    \includegraphics[width=\linewidth]{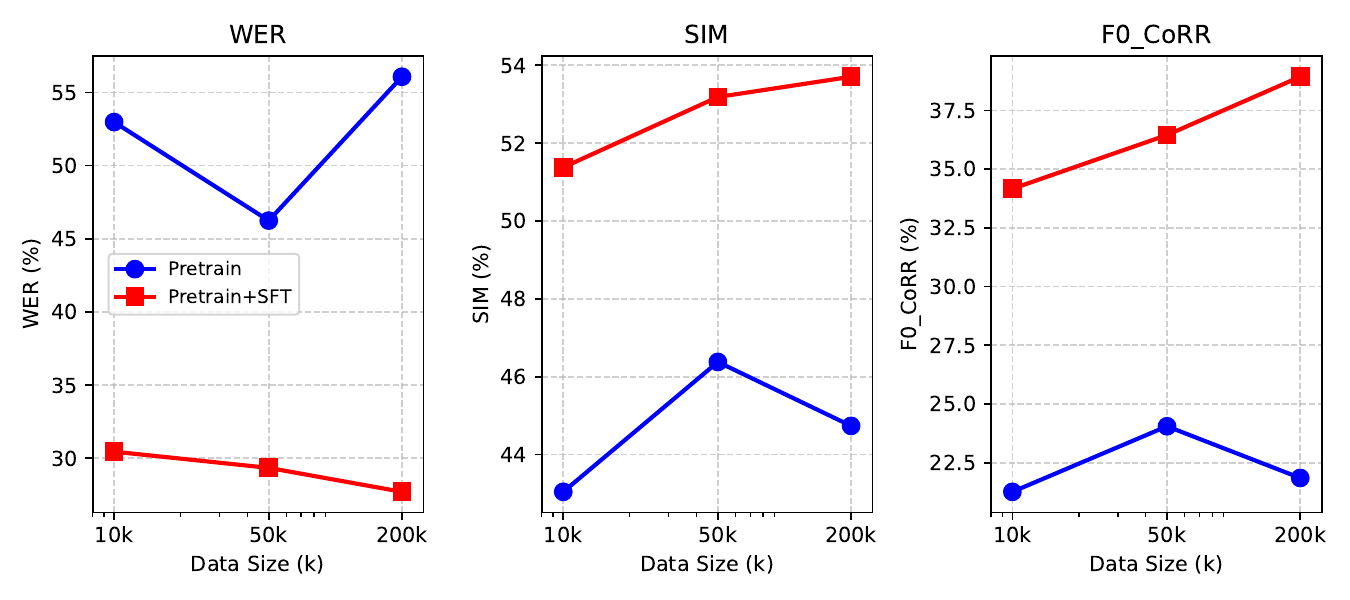}
    \caption{Scaling experiment with pseudo-parallel data. Blue curves denote models trained with pseudo-data pretraining only, while red curves denote models further fine-tuned on aligned real data after pretraining.}
    \label{fig:scaling_results}
    \vspace{-8pt}
\end{figure}

\subsection{Scaling-up Experiment (EQ3)}

We investigate the effect of scaling pseudo-parallel data generated by the N2W pipeline.
The unified tokenizer is first pretrained on different amounts of pseudo data (10k, 50k, and 200k pairs), and evaluated under two settings:

\begin{itemize}
    \item \textbf{Pretrain:} Direct evaluation after pseudo-data pretraining.
    \item \textbf{Pretrain + SFT:} Additional fine-tuning on aligned real data used in EQ2.
\end{itemize}

Figure~\ref{fig:scaling_results} shows that increasing pseudo-data scale yields limited gains under the pretrain-only setting.
However, when followed by fine-tuning on real aligned data, larger pretraining consistently leads to better performance across all metrics.
In particular, the 200k-model achieves the strongest results after fine-tuning.

These results demonstrate that large-scale pseudo pretraining provides a stronger initialization, while a small amount of real aligned data is necessary to adapt the model to the W2N task effectively.

\section{Conclusions}

We present WhispEar, a bidirectional whispered speech conversion framework based on unified semantic representations. By combining shared acoustic modeling with scalable pseudo-parallel data generation, WhispEar enables stable whisper-to-normal conversion under limited paired data and demonstrates clear gains with data scaling. We also release wEar, the largest bilingual whispered–normal parallel corpus to date.

Future work will focus on improving robustness in noisy conditions, extending multilingual generalization, and further enhancing efficiency for practical deployment.

\section{Generative AI Use Disclosure}
During the preparation of this work the authors used Gemini in order to improve language. After using this tool, the authors reviewed and edited the content as needed and take full responsibility for the content of the publication.

\bibliographystyle{IEEEtran}
\bibliography{mybib}

\end{document}